      [1999/12/01 v1.4c Il Nuovo Cimento]
\documentclass{cimento}
\usepackage{graphicx}

\title{Scalar glueball in a holographic model of QCD}%
\author{S.~Nicotri\from{a}\from{b}}
\instlist{\inst{a} Universit\`a degli Studi di Bari,
Dipartimento di Fisica \inst{b}Istituto Nazionale di Fisica
Nucleare, Sezione di Bari, Italy}

\def \de {\partial}
\def \a {\alpha}
\def \b {\beta}
\def \g {\gamma}

\def \e  {\varepsilon}

\def \L {\Lambda}

\def \m {\mu}
\def \n {\nu}

\def \r {\rho}

\def \be {\begin{equation}}
\def \ee {\end{equation}}
\def \bea {\begin{eqnarray}}
\def \eea {\end{eqnarray}}
\def \non {\nonumber}

\def \ra {\rightarrow}

\def \pr {\prime}

\def \dps {\displaystyle}
\def \fr {\displaystyle\frac}

\def\laq{~\raise 0.4ex\hbox{$<$}\kern -0.8em\lower 0.62
ex\hbox{$\sim$}~}
\def\gaq{~\raise 0.4ex\hbox{$>$}\kern -0.7em\lower 0.62
ex\hbox{$\sim$}~}

\begin{document}

\maketitle

\begin{abstract}
I describe scalar glueballs in the soft wall model of
holographic QCD (AdS/QCD).
\end{abstract}

\section{Introduction}

Recently, a lot of interest has grown around the possibility of
applying string inspired techniques to the non-perturbative
regime of QCD. The starting point is the AdS/CFT correspondence
\cite{Malda}, a conjectured duality between a maximally
supersymmetric strongly coupled conformal field theory and the
supergravity limit of type~IIB string theory, which involves
theories different from QCD. Further developments \cite{adsqcd}
have tried to apply the correspondence to QCD, induced by the
evidence of the existence of a window of energy in which QCD
shows an approximate conformal behaviour \cite{confqcd}. These
developments have taken different directions. The framework
through which I move here is the so-called soft~wall model of
AdS/QCD \cite{softwall}, a phenomenological model originally
built to holographically describe chiral symmetry breaking and
then adapted to several strong interaction processes. For a list
of other approaches the reader can refer to \cite{other}.

In the following, I discuss the scalar glueball sector and how
the spectrum and the two-point correlation function are
represented in the soft wall model. Then, I comment on the
results, comparing them with current phenomenology and lattice
data.

\section{Framework}

The considered  model is defined in a $5d$ curved space (the
bulk) with metric:
\begin{equation}
ds^2=g_{MN} dx^M
dx^N=\frac{R^2}{z^2}\,\big(\eta_{\m\n}dx^{\m}dx^{\n}+dz^2\big)
\label{metric}
\end{equation}
with $\eta_{\m\n}=\mbox{diag}(-1,+1,+1,+1)$; $R$ is the AdS
curvature radius, and the coordinate $z$  runs in the range
$0\leq z < +\infty$. QCD is supposed to live on the boundary
$z=0$, where the element $\eta_{\m\n}dx^{\m}dx^{\n}$ describes a
flat Minkowski space.

In addition to the AdS metric, the model is characterized by the
presence of a background dilaton field:
\begin{equation}
\Phi(z)=(c z)^2 \label{dilaton}
\end{equation}
exponentially coupled to the fields, whose functional form is
chosen in such a way to have linear Regge trajectories for light
vector mesons \cite{softwall}; $c$ is a dimensionful parameter
setting the scale of QCD quantities and it is of ${\cal
O}(\L_{QCD})$. It is the responsible of the breaking of
conformal symmetry and it is fixed by the experimental slope of
the rho mesons trajectory.

\section{Scalar glueballs}

$0^{++}$ glueballs can be described in QCD by the dimension four
operator ${\cal O}_S(x)=\mbox{Tr}[\b(\a_s)G^{a\,\m\n}G_{\m\n}^a]$.
In the five dimensional theory its dual field is a massless
scalar $Y(x,z)$ \cite{glueballspectrum}, whose action is given
by:
\begin{equation}\label{action}
  S=-\fr{1}{2k}\int
  d^5x\sqrt{-g}\,e^{-\phi}g^{MN}(\de_MY)(\de_NY)
\end{equation}
where $k$ is a parameter introduced to give the correct
dimension to the action. The AdS/CFT dictionary states that this
action is equivalent to the QCD partition function, in which the
source of ${\cal O}_S(x)$ is the boundary value $Y_0(x)$ of the
field $Y(x,z)$. The following relation can be written:
\begin{equation}
  Y(x,z)=\int d^4x^\pr
  K(x-x^\pr,z)Y_0(x^\pr)\;\;,
\end{equation}
where the function $K(x-x^\pr,z)$ is called bulk-to-boundary
propagator, since it links the fields in the bulk with the
sources on the boundary.

It is possible to obtain QCD correlation functions functionally
deriving the action (\ref{action}) with respect to $Y_0(x)$. The
two-point function obtained in this way is, in the limit
$q^2\ra+\infty$ \cite{glueballcorrel,forkel}:
\begin{eqnarray}\label{piads}
  \Pi_{AdS}(q^2) & = & \fr{R^3}{k}\biggl\{q^4\cdot\fr{1}{8}\left[2-2\g_E+\ln4-\ln(q^2/\n^2)\right]+\non\\
  && +q^2\left[-\fr{\n^2}{2}+\fr{c^2}{4}\left(1-4\g_E+2\ln4-2\ln(q^2/\n^2)\right)\right]+\\
  && -\fr{5c^4}{6}+\fr{2c^6}{3q^2}+{\cal
  O}\left(\fr{1}{q^4}\right)\biggr\}\;\;,\non
\end{eqnarray}
to be compared with the QCD result \cite{Paver}:
\begin{eqnarray}\label{piqcd}
  \Pi_{QCD}(q^2) & = & 2\biggl(\fr{\b_1}{\pi}\biggr)^2\biggl(\fr{\a_s}{\pi}\biggr)^2\,q^4\biggl(-\ln{\big(\fr{q^2}{\n^2}\big)}+2
  -\fr{1}{\e^{\pr}}\biggr)+4\b_1^2\,\Big(\frac{\a_{s}}{\pi}\Big)\langle{\cal
  C}_4\rangle\non\\
  &&+8\b_1^2\Big(\frac{\a_s}{\pi}\Big)^2\frac{\langle{\cal C}_6\rangle}{q^2}+{\cal
  O}\left(\fr{1}{q^4}\right)\; .
\end{eqnarray}
Matching (\ref{piads}) with (\ref{piqcd})
($\frac{R^3}{8k}=2(\b_1/\pi)^2(\a_s/\pi)^2$) one gets the fully
analytic form of the correlator. By casting it in the form:
\begin{equation}
  \Pi_{AdS}(q^2)=\dps\sum_{n=0}^\infty\fr{f_n^2}{q^2+m_n^2+i\e}
\end{equation}
it is possible to find the poles $m_n^2=4c^2(n+2)$ and the
related residues $f_n^2=\langle0|{\cal
O}_S(0)|n\rangle^2=\frac{8R^3c^6}{k}\,(n+1)(n+2)$, corresponding
to the mass spectrum and the decay constants of the scalar
glueballs. The results for the lowest lying state are:
\begin{itemize}
  \item $M_{0^{++}}=1.089$~GeV
  \item $f_{0^{++}}=\langle0|{\cal
  O}_S(0)|n=0\rangle=0.763$~GeV$^3$~~~.
\end{itemize}
The $0^{++}$ glueball turns out to be heavier than the $\r$
meson, as expected by phenomenology, but slightly lighter with
respect to the results from lattice simulations
\cite{latticeglueball}.

Another point is that in the large $q^2$ expansion there is a
dimension~two condensate, absent in QCD since there are no ways
to construct scalar local gauge invariant quantities with that
dimension \cite{zakcond2}.

Finally, the dimension four condensate $\langle{\cal
C}_4\rangle$ turns out to be negative in this picture, at odds
with the commonly used value
$\langle(\a_s/\pi)G^2\rangle\simeq0.012$~GeV$^4$
\cite{khodjamirian}.

A discussion on how to fix some of the problems exposed above
can be found in \cite{glueballcorrel, scalarmesons}.

\acknowledgments I thank P.~Colangelo, F.~De~Fazio, F.~Giannuzzi
and F.~Jugeau for collaboration.

\end{document}